# Single-Shot 60 dB Dynamic Range Laser Contrast Measurement Using Self-Referencing Spectral Interferometry

**Sasi Palaniyappan[1,a)], R.P. Johnson[1], T. Shimada[1], R.C. Shah[1], D. Jung[2], D. C. Gautier[1], B. M. Hegelich[3], and J. C. Fernandez[1]**

[1]*P-24, Plasma physics, Los Alamos National Laboratory, Los Alamos, NM. 87545, USA*

[2]*Queens University, Belfast, Ireland.*

[3]*University of Austin, Texas, USA*

[a)]Electronic email: sasi@lanl.gov

**Abstract:** High dynamic range contrast measurement of laser pulses is typically obtained by cross-correlating the laser pulse with a self-generated reference laser pulse directly in the temporal domain. Alternatively, it is also possible to measure the spectral interferogram of the laser pulse with the references pulse in the spectral domain, from which the high dynamic range contrast measurement can be extracted using inverse Fourier-transformation. Here, we demonstrate single-shot 60 dB dynamic range laser contrast measurement by inverse Fourier-transforming the measured spectral interferogram into temporal domain. The result directly provides the cross-correlation of the laser pulse with the reference pulse provided there is enough time-delay imposed between the laser and reference pulses. Then method is cross-calibrated by measuring known pre-pulses along with scanning third-order auto-correlator measurements.

## I. Introduction

Access to the regime of relativistic laser-matter interaction[1] is currently possible using petawatt and sub-petawatt class lasers providing focused-laser intensities well above $10^{20}$ W/cm$^2$. However typical laser pulses from these high power lasers contain laser pre-pulse/pedestal with laser intensity well above the laser ablation threshold of ~$10^{12}$ W/cm$^2$ extending up to several nanoseconds before the laser peak. This can significantly modify the target condition and generate plasma well before the laser peak intensity arrives at the target. This is also a major concern for the larger scale laser facilities that are currently planned and being constructed to access the extreme high field physics[2] at laser intensities above $10^{23}$ W/cm$^2$. The pre-pulse/pedestal of the laser pulse could largely arise from Amplified-Spontaneous-Emission (ASE)[3], pre-pulses from multiple round trips inside the laser cavity and post-pulses wrapping around as pre-pulses due to non-linear phase (B-integral) in the gain medium[4]. The ability to suppress the laser pre-pulses and the pedestal level below the ablation threshold is sought, to avoid generation of a long column of plasma in front of the target, which could lead to laser self focusing and beam filamentation. For this purpose, it is crucial to quantify this problem by measuring the laser contrast over a high dynamic range and large temporal window, so that we can seek an effective solution to it. Particularly for large scale laser facilities with a low repetition rate (e.g. ~1 shot/hour on the LANL Trident[5] laser facility), it is important to make that measurement in a single laser shot.

Over the years, extensive experimental effort has been devoted to single shot laser pulse characterization either directly in the temporal domain or indirectly in the spectral domain. In the temporal domain Frequency-Resolved-Optical-Gating (FROG)[6] and third order auto-correlation[7-12] are widely used for single-shot laser pulse characterization. Specifically, a single-shot third-order autocorrelator using a pulse replicator[7] has been demonstrated to measure the laser contrast with 60 dB dynamic range and 200 ps temporal window. On the other hand, numerous types of spectral interferometry such as Spectral-Shearing-Interferometry (SSI)[13,14], Spectral-Phase-Interferometry-for-Direct-Electric field-Reconstruction (SPIDER)[15] and Self-Referencing-Spectral-Interferometry (SRSI)[16-19,25,26] have also been used for single-shot laser-pulse characterization in the spectral domain. Specifically, SRSI[17] has been recently used to measure the laser pulse shape with 50 dB dynamic range within a ±0.4 ps temporal window.

In SRSI, the laser pulse to be characterized is combined with a self-created reference pulse using Cross-Polarized-Wave[20,24] (XPW) to create a spectral interferogram, which contains the laser pulse information. The spectral interferogram is then inverse Fourier-transformed into the temporal domain, where the interference term (herein called as AC term) and the non-interference tem (herein called as DC term) are isolated from each other for further analysis to extract the high dynamic range temporal laser pulse information[16,17]. In the temporal domain, the maximum achievable time-delay between the AC and DC terms is limited by the finite spectral resolution of the SRSI device, which is largely limited by the spectrometer resolution. The insufficient time-delay between the AC and DC terms leads to presence of the DC term over the entire temporal window. This limitation, in conjunction with the quality of the laser pulse contrast under study can prevent high dynamic range isolation of the AC and DC terms.

To circumvent this problem, here we propose and demonstrate that one could directly obtain the high-dynamic range cross-correlation of the laser pulse to be characterized from the inverse Fourier transformed spectral interferogram avoiding the need to isolate the AC and DC terms. Our experimental setup is mostly similar to SRSI, however the difference lies in the simpler data analysis procedure and how we create the self-referencing pulse. Figure 1 shows a schematic diagram of the idea behind the proposed method. The laser pulse and the self-generated shorter reference pulse are combined into a spectrometer with a finite time-delay producing the spectral interferogram, which is then inverse Fourier-transformed into the temporal domain. The time-domain result contains three peaks viz., an interference term (AC term), non-interference term (DC term) and a mirror replica of the AC term. Here we show that the AC term directly provides the cross-correlation of the laser pulse with the reference pulse, without the need to isolate the AC and DC terms for an iterative algorithm as originally proposed in SRSI.

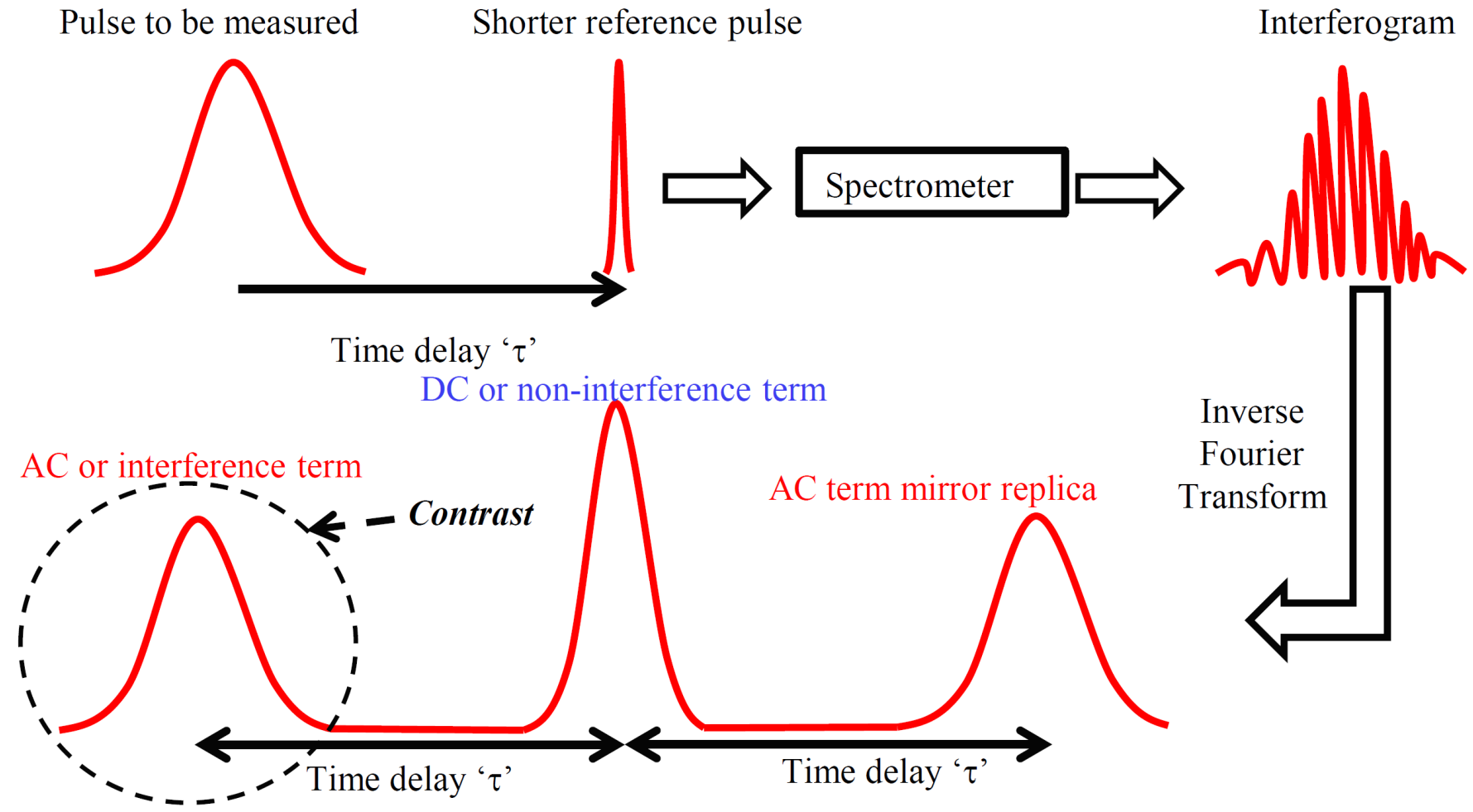

**Fig.1 Schematic of the Self-Referencing Spectral Interferometry technique**

## II. Basis of Interferometric Cross-correlation

To express the idea in quantitative terms, let $E(t)$ and $E_{ref}(t)$ be the complex-electric fields of the original input and the reference pulses respectively. The measured spectral interferogram with a given time delay τ between the input and reference pulses can be expressed as follows:

$$S(\omega)=\left|E(\omega)+E_{ref}(\omega)e^{i\omega\tau}\right|^2=\left|E(\omega)\right|^2+\left|E_{ref}(\omega)\right|^2+\left|E_{ref}(\omega)\right|\left|E(\omega)\right|e^{-i\Delta\phi(\omega)}e^{-i\omega\tau}+\left|E_{ref}(\omega)\right|\left|E(\omega)\right|e^{i\Delta\phi(\omega)}e^{i\omega\tau} \quad (1)$$

where $\Delta\phi=\phi(\omega)-\phi_{ref}(\omega)$ is the spectral phase difference between the input and reference pulses. If the reference pulse is transform-limited with flat spectral phase, the measured spectral phase difference essentially provides the input pulse spectral phase i.e., $\Delta\phi(\omega)=\phi(\omega)-\phi_{ref}(\omega)\equiv\phi(\omega)$. With this, the equation (1) can be now rewritten as

$$S(\omega)=\left|E(\omega)\right|^2+\left|E_{ref}(\omega)\right|^2+\left|E(\omega)\right|\left|E_{ref}(\omega)\right|e^{-i\phi(\omega)}e^{-i\omega\tau}+\left|E(\omega)\right|\left|E_{ref}(\omega)\right|e^{i\phi(\omega)}e^{i\omega\tau} \quad (2)$$

$$S(\omega)=\left|E(\omega)\right|^2+\left|E_{ref}(\omega)\right|^2+E^*(\omega)\left|E_{ref}(\omega)\right|e^{-i\omega\tau}+E(\omega)\left|E_{ref}(\omega)\right|e^{i\omega\tau} \quad (3)$$

The equation (3) contains four terms in it viz., the input pulse spectral intensity, the reference pulse spectral intensity, the point-wise spectral product of the input and reference pulses multiplied by a linear spectral phase and its complex conjugate. A direct inverse Fourier-transformation of the measured spectral interferogram into temporal domain yields:

$$S(t) = E^*(-t) \otimes E(t) + E^*_{ref}(-t) \otimes E_{ref}(t) \; + F(t-\tau) + F^*(-t+\tau) \qquad (4).$$

where $F = E^*(-t) \otimes \left| E_{ref}(t) \right|$.

In the equation (4), the first two terms correspond to field autocorrelation functions of the input and the reference pulses respectively, centered at zero time delay. This is known as the DC or the non-interference term, which originates from the first two terms in equation (3). The third and fourth terms of equation (4) are the cross-correlation of the input pulse with the reference pulse centered at time delays -τ and τ respectively, known as the AC or the interference terms. The AC terms represent the cross-correlation of the uncharacterized input with the reference pulse. In our setup the reference pulse amplitude profile is cube of the input pulse amplitude profile due to the non-linear optical process. Hence the AC terms are in fact cross correlation of the input pulse. The AC terms originate from the third and fourth terms in equation (3). Here, we select the 1st AC term appearing at an earlier time delay '-τ' as the cross-correlation measurement of the input pulse with high dynamic range. Since the AC terms are mirror replicas of each other, the 2nd AC term appearing at a later time delay 'τ' can also be equivalently considered to represent the laser pulse contrast measurement without any loss of generality. In practice, the DC term could be present over the entire temporal window due to finite time delay of 'τ', depending on the nature of the laser pulse being characterized. Note that the AC term provides only the field cross-correlation of the input and the reference pulses, which should be squared to obtain their intensity cross-correlation, which is typically known as the laser contrast.

**III. Experimental Setup**

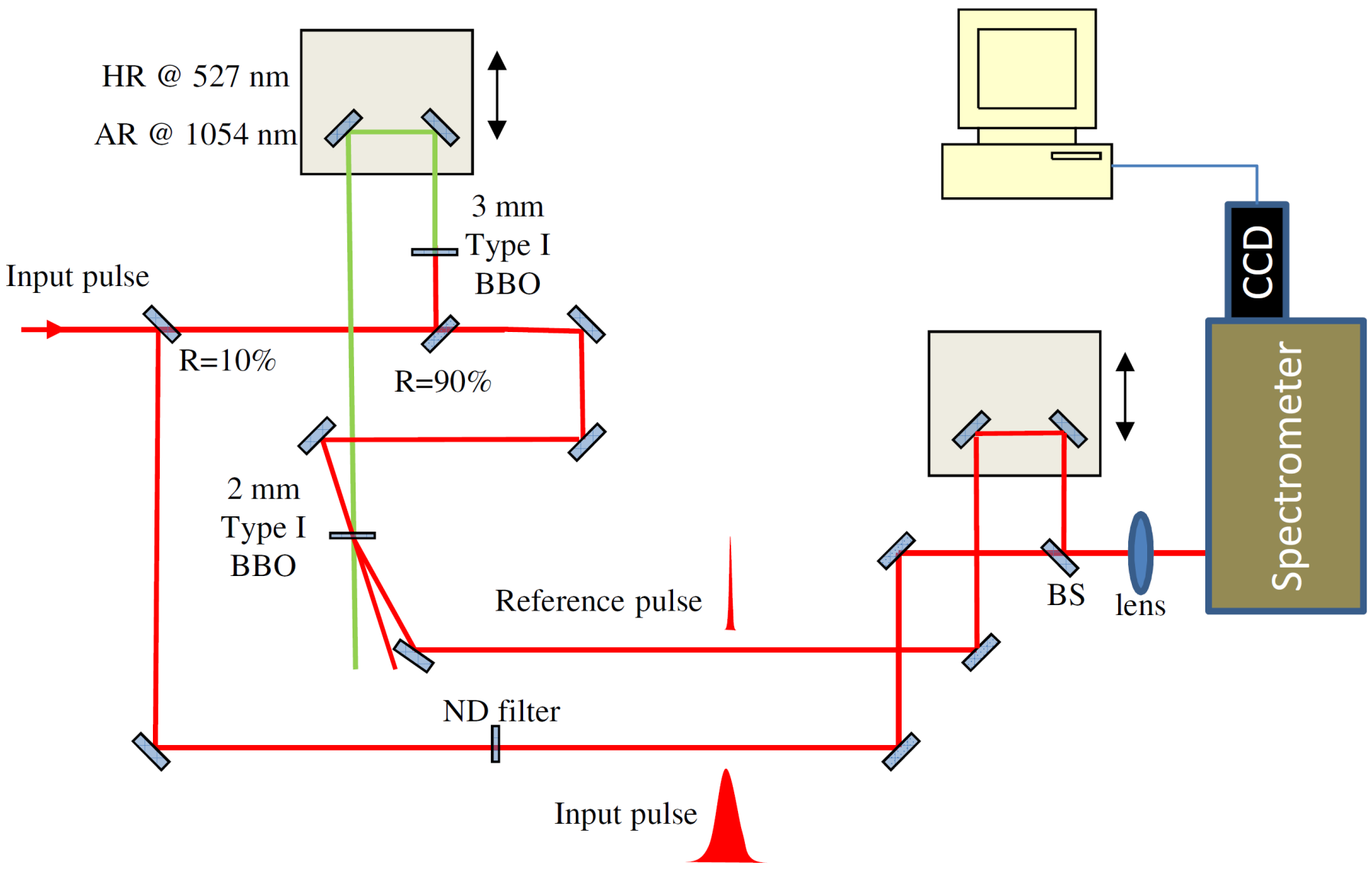

**Fig. 2 schematic of the self-referencing spectral interferometric cross-correlator experimental setup implemented at Trident laser facility.**

Figure 2 shows the schematic of the interferometric cross-correlator implemented at the front-end of the Trident laser facility at the Los Alamos National Laboratory. The input pulse to be characterized (200 μJ, 1054 nm, ~500 fs, 5 Hz, s-polarized) is initially split with a 10/90 beam splitter. The 90% transmitted pulse is then used to create a reference pulse via the low-gain OPA technique[21] as follows. In the transmitted beam, another 90/10 beam splitter is used to transmit the 10% of the beam while 90% of the beam is reflected and consecutively frequency doubled in a 3 mm thick type I BBO crystal with 150 μJ output at 527 nm. The residual collinear input pulse at 1054 nm from the type I BBO crystal is further strongly attenuated by two consecutive dichroic mirrors with high reflectance at 527 nm and anti-reflection coated at 1054 nm. The frequency doubled pulse (i.e., the pump) following that same collinear path is then combined with the 10% transmitted pulse from the 2$^{nd}$ beam splitter (i.e., the signal) with zero time delay between them in a 2 mm thick type I BBO crystal in the low gain regime in a non-collinear configuration to amplify the signal pulse via Optical-Parametric-Amplification (OPA). This in turn produces a residual idler pulse at same wavelength as the signal pulse at 1054 nm, which is used as the reference pulse for the measurement. The reference pulse has much shorter pulse duration and higher contrast than the input pulse, because its intensity profile is cube of the original input pulse intensity profile[21]. In fact the low-gain OPA technique was originally implemented at the Trident laser facility to enhance the laser contrast by cubing the input laser intensity profile[21]. Our experimental setup is similar to the SRSI except the reference pulse in SRSI is self-generated using a fraction of the laser pulse to be characterized via Cross-Polarized-Wave (XPW)[20] technique in a single non-linear crystal in a collinear geometry.

The reference pulse with 30 μJ is easily isolated from the pump and signal pulses due to the non-collinear configuration of the type I BBO crystal. The isolated reference pulse and the original 10% reflected input pulse from the 1st 10/90 beam splitter are then collinearly combined in a 50/50 beam splitter with an adjustable time delay between them using a linear translation stage. Note that the energies of the laser pulses during the low-gain OPA process mentioned here represent the typical numbers. Finally, the collinearly combined input and reference pulses with a finite time-delay between them are then simply focused at the entrance slit of an imaging spectrometer (Bruker Optics, Surespectrum 500is with 1200 lines/mm grating blazed at 500 nm) with 10 μm slit width using an f/10 focusing lens. The f/# of the final focusing lens is chosen to be comparable to the f/# of the spectrometer (f/8) to obtain best possible spectral resolution. The spectrometer provides a spectral dispersion of 1.6 nm/mm with spectral resolution of 0.01 nm for a slit width of 10 μm. The spectral interferogram is captured with a -15°C cooled silicon CCD camera (Apogee Alta® U8300) with a 2D-array of 3326 X 2504 pixels (pixel size of 5.4 μm X 5.4 μm). For the above mentioned imaging spectrometer, the CCD spectral calibration yields 0.008 nm spectral width per pixel which is close to the spectrometer resolution of 0.01 nm for 10 μm slit width.

**IV. Single-shot contrast measurement at the Trident laser facility front-end**

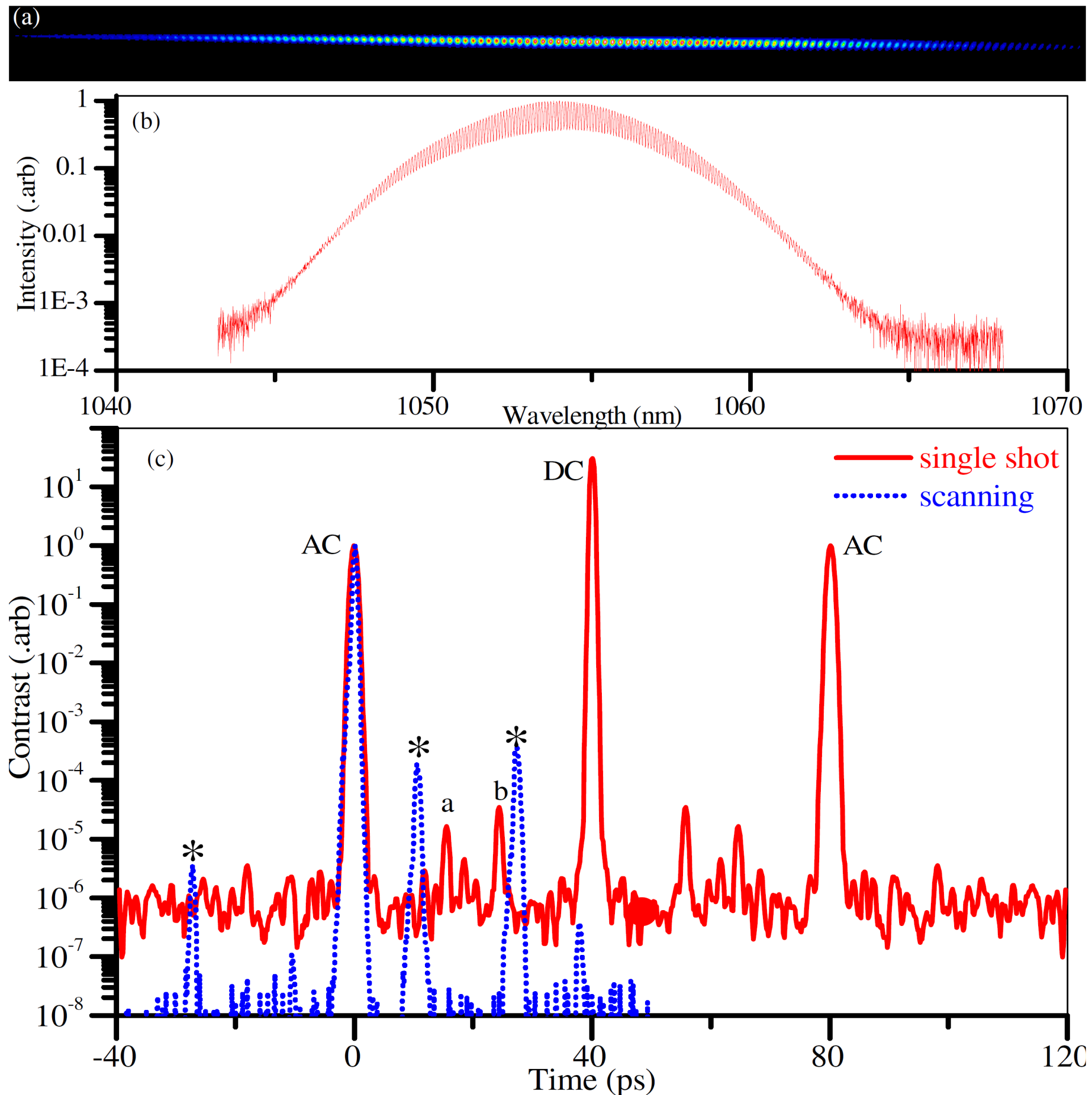

**Fig. 3 (a) measured interferogram of laser and reference pulses, (b) line-out of the spectral interferogram and (c) single-shot contrast measurement (AC peak in the red solid line) along with scanning third-order auto-correlator measurement (blue dotted line - * indicates known artifacts in the scanning third-order auto-correlator measurement).**

Figure 3(a&b) show the CCD image and the line-out of measured spectral interferogram of the laser pulse with the self-generated reference pulse with 42 ps time-delay between them respectively. The interferogram has a dynamic range of ~35 dB with 24.5 nm wavelength range (from 1043.3 nm to 1067.8 nm) and fringe separation of ~0.09 nm. The solid red line in figure 3(c) shows the result obtained by inverse Fourier-transforming the spatially averaged spectral interferogram and squaring it. The single-shot laser contrast measurement (marked as AC peak) is consistent with independent scanning third-order auto-correlator (Rincon, Delmar Photonics) measurement (dotted blue line).

## V. Calibration

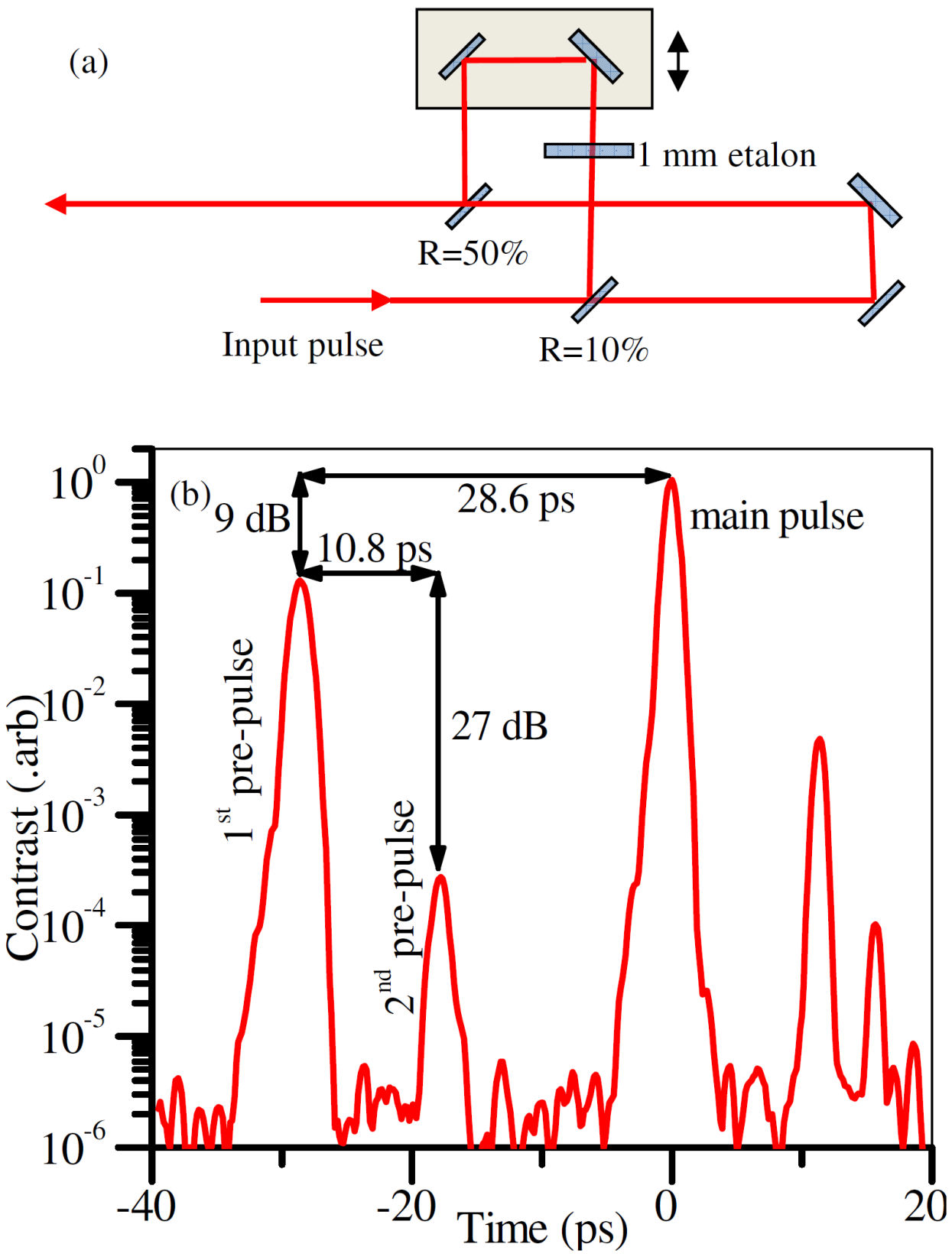

**Fig. 4 (a) contraption used to make known pre-pulses at the trident laser front-end and (b) measurement of the imposed pre-pulses using the interferometric cross-correlator.**

Figure 4(a) shows the contraption used to impose known pre-pulses to the laser pulse at the Trident front-end. A 10/90 beam splitter sends the 10% reflected input pulse into a 1 mm thick etalon, which is then combined with the remaining 90% transmitted light using a 50/50 beam splitter via an adjustable path-delay. Over all, the pre-pulse injection arm is 8.6 mm shorter than the main pulse arm. This means that the first pre-pulse will be 28.7 ps ahead of the main pulse with the peak 9.5 dB lower than the input pulse (50% of the 90% transmitted pulse Vs. 50% of the 10% reflected pulse). The second reflection from the etalon will be 10.3 ps later with the peak 28 dB lower than the first reflection (4% of 4% - double reflection within the etalon). The measurement of the imposed pre-pulse using interferometric cross-correlator in fig. 4(b) shows pre-pulses that are consistent with the expected results.

**VI. Single-shot contrast measurement of Trident full energy system shot using interferometric cross-correlator**

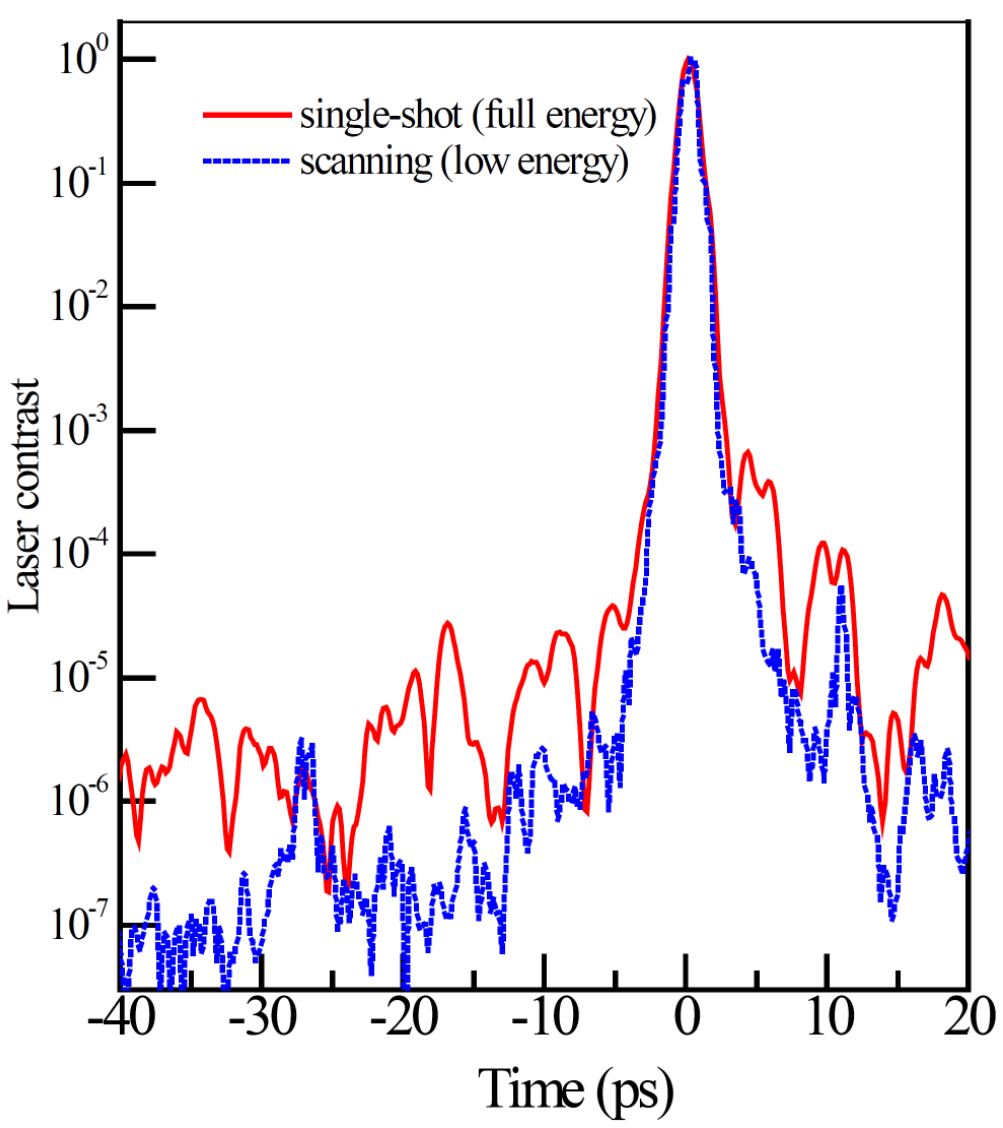

**Fig. 5 single shot interferometric cross-correlator full energy (solid redline) and scanning third-order low-energy (blue dotted line) contrast measurements of the Trident laser pulse at the target area.**

Figure 5 shows the single-shot Trident full system shot (80 J, 650 fs with one-shot per hour) contrast measurement using interferometric cross-correlator (red solid line) along with the scanning third-order measurement (dotted blue line) obtained by propagating few mJ laser pulses from the trident front-end along the laser amplification chain with 5 Hz repletion rate. Details of laser sampling during the full system shot has been described elsewhere[22,23]. The result shoes that the interferometric cross-correlation measurement is consistent with the scanning measurement. However, the interferometric cross-correlator measurement is only up to ~50 dB dynamic range. This is limited by the current laser sampling system which allows only ~ 100 μJ at the spatial filter due to air breakdown. This can be further improved by designing a vacuum spatial filter.

## V. Conclusion

In conclusion, we have demonstrated 60 dB single-shot contrast measurement using interferometric cross-correlation at the Trident laser facility. The device is calibrated by measuring known imposed pre-pulses and with scanning third-order auto-correlator measurements. We have also used the device to measure the laser contrast with Trident full system shot. To our knowledge, this is the first single-shot contrast measurement on a high power laser full system shot. The dynamic range of the measurement with full system shot can be further improved by designing a vacuum spatial filter which will allow us to collect more laser energy. Also, a CCD with better signal to noise ratio with smaller pixel size can increase the overall dynamic range of the measurement.

## Acknowledgments

S. P. acknowledges the support of S. Batha. The research was supported by LANL Inertial Confinement Fusion (ICF) program, funded by the U.S. Department of Energy (DOE) office.